\documentclass[%
 reprint,
 amsmath,amssymb,
 aps,
]{revtex4-2}

\usepackage{amsmath,amssymb,amsbsy,amstext, amsthm, simplewick, amsfonts}

\usepackage{graphicx}

\usepackage{siunitx}
\usepackage{upgreek}
\usepackage{framed}
\usepackage{wrapfig}
\usepackage{multirow}
\usepackage{bbm}

\usepackage[absolute,overlay]{textpos}

\begin{document}

\begin{textblock*}{5cm}(18.46cm,1cm)  
    \small YITP-25-67
    
\end{textblock*}

\preprint{APS/123-QED}


\title{The Dual Primordial Black Hole Formation Scenario}

\author{Xinpeng Wang$^{1,2,3}$}
\email{xinpeng.wang{}@ipmu.jp}
 \author{Misao Sasaki$^{2,4,5,6}$}%
 \email{misao.sasaki{}@ipmu.jp}
\author{Ying-li Zhang$^{1,7,8}$}%
 \email{yingli{}@tongji.edu.cn}
 
 \affiliation{$^1$ School of Physics Science and Engineering, Tongji University, Shanghai 200092, China\\
$^2$Kavli Institute for the Physics and Mathematics of the Universe (WPI), The University of Tokyo Institutes for Advanced Study, The University of Tokyo, Chiba 277-8583, Japan\\
$^3$ Department of Physics, Graduate School of Science, The University of Tokyo, Tokyo 113-0033, Japan\\
$^4$Center for Gravitational Physics and Quantum Information,
Yukawa Institute for Theoretical Physics, Kyoto University, Kyoto 606-8502, Japan\\
$^5$Leung Center for Cosmology and Particle Astrophysics,
National Taiwan University, Taipei 10617, Taiwan\\
$^6$Asia Pacific Center for Theoretical Physics, Pohang 37673, Korea\\
$^7$Institute for Advanced Study of Tongji University, Shanghai 200092, China\\
$^8$Center for Gravitation and Cosmology, Yangzhou University, Yangzhou 225009, China}




\date{\today}

\begin{abstract}
We report a novel mechanism where two families of primordial black holes (PBHs) may form at nearly the same comoving scales but at two different epochs. It is realized in two-stage inflation where a non-inflationary stage is sandwiched by the two inflationary stages.
In this case, smaller PBHs form when the comoving scale of interest re-enters the horizon during the break period, and larger PBHs form when the scale re-enters the horizon after inflation. 
This mechanism may realize both reheating of the universe through the evaporation of ultralight PBHs formed during the break stage and the dark matter by those formed after inflation. 
We show that this scenario may give rise to a distinctive signature in the stochastic gravitational wave background that can be tested by the near-future gravitational wave observatories such as LISA and DECIGO. Our work thus provides a unified observational window into the physics of inflation, reheating, and dark matter.
\end{abstract}

\maketitle


\section{\label{sec:level1}Introduction}

Primordial Black Holes (PBHs), formed from rare but significantly amplified curvature perturbations from inflation, are intriguing objects that encapsulate rich physical phenomena~\cite{zel1966hypothesis, Hawking:1971ei, Carr:1974nx, Meszaros:1974tb, Khlopov:1985jw}. 
They are a natural candidate for cold dark matter (CDM) \cite{Chapline:1975ojl} since they interact only gravitationally. 
Specifically, PBHs with asteroid masses in the range $10^{17} {\rm g}\sim 10^{21} \rm g$ may constitute all of the CDM \cite{Niikura:2019kqi, Katz:2018zrn, Smyth:2019whb, Montero-Camacho:2019jte}.
Furthermore, ultra-light PBHs with masses smaller than $5\times 10^{8}\rm g$ evaporate before big-bang nucleosynthesis (BBN). 
These PBHs may induce an early dark-matter-dominated era, heat the universe via Hawking radiation\cite{Hidalgo:2011fj,Carr:1976zz}, and leave detectable imprints in the gravitational wave background \cite{Papanikolaou:2020qtd, Domenech:2020ssp, Domenech:2023jve,Martin:2019nuw,Domenech:2024wao, del-Corral:2025fca}.

The conventional scenario is that a rare, large amplitude curvature perturbation gravitationally collapses to form a PBH when the scale enters the Hubble horizon during the radiation-dominated stage. 
Such large curvature perturbations are considered to be seeded by quantum vacuum fluctuations during inflation.
Although the amplitude of the curvature perturbations on large scales ($\gtrsim 1$ Mpc), which exit the Hubble horizon $50\sim60$ e-folds before the end of inflation, is strongly constrained by cosmic microwave background and large scale structure observations, those on small scales ($\ll 1$ Mpc) are poorly constrained due to the lack of observational data that contain primordial signals.
This gives rise to various possibilities of phenomena on these small scales. 



In this letter, we explore the dual PBH formation scenario in the two-stage model of inflation in which the inflationary expansion temporarily halts during inflation \cite{Polarski:1995zn,Zelnikov:1991nv,Allahverdi:2007ts,Namjoo:2012xs,Kohri:2012yw,Pi:2019ihn,Wang:2024vfv,Kim:2025dyi}. 
This divides inflation to three stages: the first slow-roll stage, the intermediate decelerated expansion stage, and the second slow-roll stage. 
During the intermediate stage, a certain range of comoving wavenumbers re-enters the horizon. Hence, the perturbations in this range may collapse to form PBHs if the amplitude is large enough. 
The same range of comoving scales exits the horizon during the second stage of inflation, and re-enters the horizon again at a much later epoch after inflation.
Some of the perturbations in this comoving range may collapse to PBHs at that epoch.
Since the mass of PBHs is inversely proportional to the Hubble parameter at the time of horizon re-entry, this process can lead to the formation of ultra-light PBHs from the first re-entry during inflation and much heavier PBHs from the second (regular) re-entry after inflation.

The ultra-light PBHs with masses $\lesssim 10^9\rm g$ if abundant enough may dominate the universe after inflation to realize an early matter-dominated stage, heating the universe before Big Bang Nucleosynthesis (BBN) through Hawking evaporation \cite{Carr:1976zz,Hawking:1975vcx,Hidalgo:2011fj,RiajulHaque:2023cqe}. 
These PBHs evaporate almost instantaneously, leading to a rapid transition from an early matter-dominated era to a radiation-dominated era. 
This rapid transition can induce significant inhomogeneities and generate observable gravitational waves (GWs) within a frequency range accessible to next-generation space-based GW observatories such as LISA, Taiji, TianQin, and DECIGO \cite{LISA:2022yao,TianQin:2015yph,Ruan:2018tsw,Hu:2017mde,Kawamura:2011zz,Kawamura:2006up,Kawamura:2020pcg,TianQin:2020hid}.
The heavier PBHs formed at a later time are massive enough ($\gtrsim 10^{18}\rm g$) to survive until today and may constitute CDM. 
The associated induced GWs could be detected by ground-based GW observatories like LIGO/Virgo/KAGRA \cite{LIGOScientific:2018mvr, LIGOScientific:2020ibl, LIGOScientific:2021djp,KAGRA:2013rdx,VIRGO:2014yos,Smith:2002dz,Somiya:2011np,Aso:2013eba} and ET \cite{Punturo:2010zz,Hild:2010id} in the future.
We find that the unique GW signals are generated in this scenario, and they may be used to probe the evolution of the primordial universe.

\section{\label{sec:level1} Dual PBHs from two-stage inflation with a break}

\begin{figure}[htbp]
\includegraphics[width=.48\textwidth]{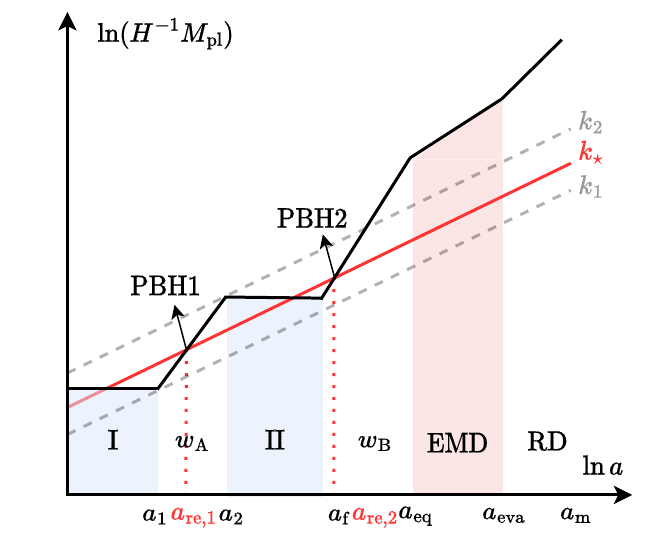}
\caption{\label{fig1}
The space-time diagram of our scenario. The two-stage inflation (denoted by ``I" and ``II") is intermediated by a break stage with EOS $w_A$. After the end of inflation, the universe enters the reheating stage with EOS $w_B$, followed by a PBH-dominated stage (EMD: early matter dominance) before the radiation-dominated stage (RD). 
The black solid line shows the evolution of the Hubble horizon. 
The red solid line represents the mode of primordial curvature perturbation, the comoving wavelength $k_\star$, which exits the Hubble horizon twice during two inflationary stages. Correspondingly, two sets of PBHs are formed at the twice re-enteries into the Hubble horizon, denoted as ``PBH1" and ``PBH2", respectively.}
\end{figure}

We consider a two-stage inflationary scenario (denoted by stages I and II in the following) with an intermediate break stage, as illustrated in Fig.~\ref{fig1}. 
We approximate the Hubble parameters of stages I and II to be constant, given by $H_{\rm{I}}$ and $H_{\rm{II}}$, and  
denote the end of stages I and II by $t_{\rm 1}$ and $t_{\rm f}$, respectively, and the beginning of stage II by $t_2$.
During the break stage $t_1<t<t_2$, the equation of state (EOS) parameter is $w_A$.
After the end of inflation $t>t_{\rm f}$ the EOS is $w_B$. 
To assure the decelerated expansion, we assume both $w_A$ and $w_B$ are greater than $-1/3$.
We consider the scenario where two sets of PBHs (denoted by PBH1 and PBH2) are seeded by curvature perturbations with the comoving wavenumber $k_\star$, which exits the Hubble horizon twice during the two inflationary stages, and reenters the horizon twice during $w_A$ and $w_B$ dominated stages.


\subsection{PBH formation}

During the break stage, the energy density decreases as $\rho_A\propto a^{-2n_A}$ ($n_A\equiv3(1+w_A)/2$) and the Hubble parameter is given by $H=(n_At)^{-1}$.

The PBH mass at the formation time is given by 
\begin{align}
    M_{\rm PBH}(t_{\rm re})=\frac{4\pi\gamma M_{\rm pl}^2}{ H(t_{\rm re})},
    \label{mpbhf}
\end{align}
where $t_{\rm{re}}$ is the horizon reentry time, $\gamma$ is a constant of order unity ($\gamma\approx0.2$ if PBHs form during radiation dominance) and $M_{\rm pl}=\sqrt{\hbar c/(8\pi G)}\approx 4.3\times10^{-6}\rm g$. Below, we ignore the equation of state dependence of $\gamma$ and assume $\gamma\approx0.2$ for simplicity.
The formation time of PBH can be estimated by the free fall time of the Hubble patch $t_{\rm ff}\sim 1/H$, which implies that the duration of the intermediate stage should be longer than $\sim 1$ e-fold. 

Since $H_{\rm II}<H_{\mathrm{re,1}}<H_{\rm I}$, the mass of PBH1 at formation will be in the range $M_{1,\rm max}>M_1(t_{\rm re,1})>M_{1,\rm min}$. 
Let us fix the Hubble parameter of the first stage to $H_{\mathrm{I}}=10^{-5}M_{\rm pl}$. This gives the minimum value of PBH1 mass $M_{1,\rm min}\sim1$g.
Since $H_{\rm{II}}=\exp[-3\Delta N(1+w_A)/2]H_{\rm{I}}$ where $\Delta N$ is the number of e-folds of the break stage, 
the maximum value of PBH1 mass $M_{1,\rm max}\sim10^8\rm g$ for $\Delta N\approx12/(1+w_A)$ .

\begin{figure}[htbp]
\includegraphics[width=.48
\textwidth]{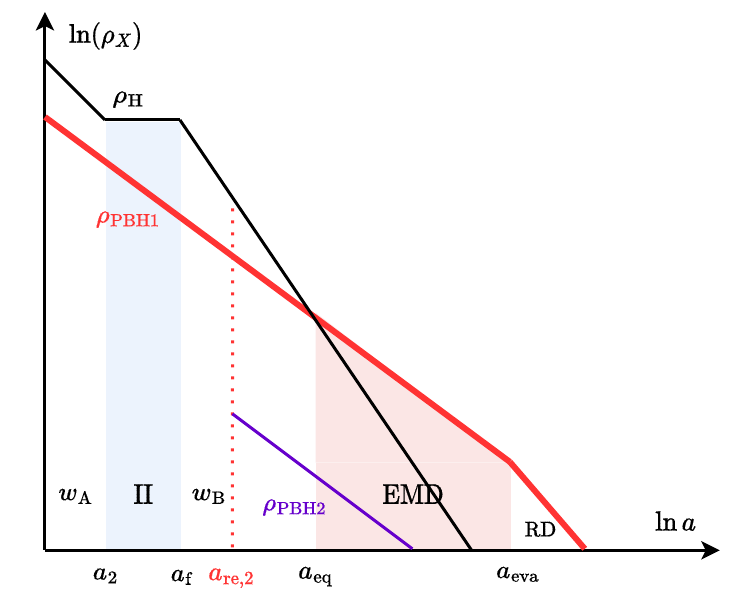}
\caption{\label{rhopbh}An illustration of the evolution of energy density.
The black solid line shows the evolution of the energy density without PBHs. 
The red and purple solid lines show the energy densities of PBH1 and PBH2, respectively.}
\end{figure}

For simplicity, let us focus on the case $w_B>0$. 
Then the energy density of the universe right after inflation decays faster than that of matter, $\rho_{\rm B}\propto a^{-2n_B}$ $ (2n_B\equiv 3(1+w_{B})>3)$.
Correspondingly, the universe is dominated by PBH1 after $t_{\rm{eq}}$, as illustrated in Fig.~\ref{rhopbh}.
Denoting $a_{\rm eq}$ as the scale factor when $\rho_{\mathrm{PBH1}}=\rho_{\mathrm{B}}$, 
we obtain the following relation,
\begin{equation}
    \begin{aligned}
   \rho_{\rm{H}}(t_{\rm re,1}) &\left(\frac{a_{2}}{a_{\rm{re,1}}}\right)^{-2n_A}\left(\frac{a_{\rm eq}}{a_{\rm f}}\right)^{-2n_B}\\&= \rho_{\rm{PBH,1}}(t_{\rm re,1})\left(\frac{a_{\rm eq}}{a_{\rm{re,1}}}\right)^{-3},
\end{aligned}
\end{equation}
where $\rho_{\rm{H}}(t_{\rm re,1})$ is the energy density of the universe at $t_{\rm re,1}$, $a_{\rm f}$ is the scale factor at the end of inflation.
Then the abundance of PBH1 at the formation reads
\begin{equation}
    \begin{aligned}
    \beta_{1}&\equiv\frac{\rho_{\rm{PBH1}}(t_{\rm re,1})}{   \rho_{\rm{H}}(t_{\rm re,1})}\\
    &=\left[\left(\frac{a_{2}}{a_{\rm{re,1}}}\right)^{n_A}\left(\frac{a_{\rm eq}}{a_{\rm f}}\right)^{n_B}\right]^{-2}\left(\frac{a_{\rm eq}}{a_{\rm{re,1}}}\right)^{3}.
    \label{abund1}
\end{aligned}
\end{equation}
Since PBH1 should not dominate the universe before stage II, this imposes a constraint on $\beta_1$ for $w_A>0$ such that
\begin{align}
    \beta_{1}\left(\frac{a_{2}}{a_{\rm re,1}}\right)^{2n_A-3}<1.
\end{align}

During the $w_B$-dominated stage, $k_\star$ re-enters the horizon at $t_{\rm re,2}$ \footnotemark[2]\footnotetext[2]{It can be proved that it is impossible for those modes to re-enter the horizon during or after the early PBH dominated era. See a proof in Appendix A.}. 
Since $ k_\star=a_{\mathrm{re,1}}H_{\mathrm{re,1}}=a_{\mathrm{re,2}}H_{\mathrm{re,2}}$, we obtain the relation 
\begin{align}
   \frac{a_{\mathrm{re,2}}}{a_\mathrm{re,1}}=\frac{H_{\mathrm{re,1}}}{H_\mathrm{re,2}}=\left(\frac{a_{2}}{a_{\mathrm{re,1}}}\right)^{n_A}\left(\frac{a_\mathrm{re,2}}{a_{\mathrm{f}}}\right)^{n_B}\,.
   \label{hubeq}
\end{align}
Denoting $M_2$ as the mass of PBH2, using \eqref{hubeq} and  \eqref{abund1}, we have
\begin{align}
    \frac{M_{2}}{M_{1}}=\frac{a_{\rm re,2}}{a_{\rm re,1}}=\left(\beta_{1}\right)^{1/(2n_B-2)}\left(\frac{a_{\rm eq}}{a_{\rm re,1}}\right)^{(2n_B-3)/(2n_B-2)}\,.
    \label{massr}
\end{align}

\subsection{Reheating from evaporation of PBH1}
As explained in the previous subsection, during the $w_B$-dominated stage, since $w_B>0$, the energy density of $w_B$ decays more quickly than that of PBH1, so the ratio $\rho_{\rm PBH1}/\rho_{\rm B}$ increases monotonically and finally becomes greater than unity after $t_{\rm eq}$, leading to a PBH-dominated universe. For PBH1 with small mass at formation $M_1(t_{\rm re,1})\lesssim10^8\rm g$ (as we will see later in the constraint on $M_1$), they will evaporate quickly due to Hawking radiation. Denoting $t_{\rm eva}\equiv M_{\rm 1}^3(t_{\rm re,1})/(3\alpha M_{\rm pl}^4)$ as the evaporation time of PBH1 with $ \alpha\approx0.855$~\footnotemark[1]\footnotetext[1]{Since we are considering PBH1 whose mass $M_1\ll 10^{11}\rm g$, the spin-weighted degree of freedom $g_{H}(T_{\rm PBH})\approx108$\cite{Husdal:2016haj, Saikawa:2018rcs} is used in the following calculation. 
Here $T_{\rm PBH}\equiv M_{\rm pl}^2/ M_{1}$, then $\alpha\approx 3.8\pi g_{H}(T_{\rm PBH})/480$}, during PBH-dominated stage, PBH1 will evaporate away so that the universe at last enters the radiation-dominated stage after $t_{\rm eva}$.

On the other hand, to maintain a successful nucleosynthesis, PBH1 should evaporate at temperature $T_{\rm eva}>4\rm MeV$. Since
\begin{align}\label{tempeva}
T_{\rm{eva}}\approx2.8\times10^{10}\mathrm{GeV}\left(\frac{M_{1}}{1\rm {g}}\right)^{-3/2}\left(\frac{g_{s\star}(T_{\rm{eva}})}{106.75}\right)^{-1/4},
\end{align}
this implies $M_{1}\lesssim 5\times 10^{8}\rm g$, which is consistent with the assumption $\Delta N\approx12/(1+w_A)$ stated in the previous subsection.

Moreover, a necessary condition for our scenario is that the lifetime of PBH1 is long enough so that they evaporate after their domination, i.e. $t_{\rm eva}>t_{\rm eq}$, or equivalently ${a_{\rm eva}}/{a_{\rm eq}}>1$. Considering $H_{\rm eva}\ll H_{\rm eq}$ and $H_{\rm re,1}\ll H_{\rm 1}$, by integrating  ${\rm d}t={\rm d}\ln a/H$, we obtain 
\begin{align}\label{aratio1}
&\left(\frac{a_{\rm eva}}{a_{\rm re,1}}\right)^{3/2}\approx(\beta_{1})^{1/2}\frac{3}{2n_A}\frac{t_{\rm eva}}{t_{\rm re,1}}.
\end{align}
Inserting \eqref{mpbhf} and \eqref{massr} into \eqref{aratio1}, we obtain the ratio ${a_{\rm eva}}/{a_{\rm eq}}$ as a function of masses of PBHs and the initial abundance of PBH1 $\beta_{1}$,
\begin{eqnarray}
&\dfrac{a_{\rm eva}}{a_{\rm eq}}=\dfrac{a_{\rm eva}/a_{\rm re,1}}{a_{\rm eq}/a_{\rm re,1}}
\approx\left(\dfrac{2\pi\gamma }{\alpha}\right)^{2/3}(\beta_{1})^{2n_B/(6n_B-9)}
\nonumber\\
&\quad\times\left(\dfrac{M_2}{M_1}\right)^{(2-2n_B)/(2n_B-3)}\left(\dfrac{M_{1}}{M_{\rm pl}}\right)^{4/3}.
\end{eqnarray}
Therefore, the condition ${a_{\rm eva}}/{a_{\rm eq}}>1$ imposes a constraint on $\beta_{1}$ such that
\begin{equation}
    \begin{aligned}
&\text{(Condition 1)}\\&\beta_{1}>\left(\frac{1280\pi\gamma }{\alpha}\right)^{3/n_B-2}\left (\frac{M_{2}}{M_{1}}\right)^{3-3/n_B}\left(\frac{M_{1}}{M_{\rm pl}}\right)^{6/n_B-4}.
    \label{con1}
\end{aligned}
\end{equation}
Combining with the constraint $\beta_1<1$, considering $M_{1}\lesssim 5\times 10^{8}\rm g$, eq.~\eqref{con1} gives a constraint on $M_1$ and $M_2$ when $w_B$ is fixed. For instance, if $w_B=1/3$ ($n_B=2$), eq.~\eqref{con1} becomes
\begin{align}
        1>\beta_{1}>{4.43\times 10^{-2}}\left (\frac{M_{2}}{10^{17}\rm g}\right)^{3/2}\left(\frac{M_{1}}{10^8 \rm g}\right)^{-5/2},
    \label{betacondm}
\end{align}
which constrains $M_2\lesssim10^{19}{\rm g}$ for $M_1<5\times 10^8 {\rm g}$, implying that PBH2 may account for the cold dark matter (CDM). We note that the constraint on $M_2$ can be relieved for $w_B>1/3$. For example, in case of kination where $w_B=1$, ~\eqref{con1} gives $M_2\lesssim10^{23}{\rm g}$.

Besides \eqref{con1}, another necessary condition for our scenario is that, in order not to spoil the standard cosmology, PBH2 should not dominate the universe before the radiation-matter equality $t_{\rm m}$, i.e. $\rho_{\rm PBH,2}(t_{\rm m})/\rho_{\rm r}(t_{\rm m})< 1$. This implies
\begin{align}\label{beta2con}
    \beta_2 \left(\frac{a_{\rm eq}}{a_{\rm re,2}}\right)^{2n_B-3}\xi\lesssim1,
\end{align}
where $\beta_2\equiv\rho_{\rm PBH2}(t_{\rm re,2})/\rho_{\rm H}(t_{\rm re,2})$ is the abundance of PBH2 at the formation, $\xi\equiv{a_{\rm m}}/{a_{\rm eva}}$ is the total expansion of the universe from the radiation-dominated era. Inserting \eqref{massr} into \eqref{beta2con}, we obtain a constraint on $\beta_{2}$
\begin{equation}
   \begin{aligned}
  &\text{(Condition 2)}\\
    &\frac{\beta_{2}}{\beta_{1}}\lesssim\frac{1}{\xi}\frac{M_{1}}{ M_{2}}\approx
    \left(\frac{t_{\rm eva}}{t_{\rm m}}\right)^{1/2}\frac{M_{1}}{ M_{2}}\\
    &\quad\lesssim 6\times10^{-17}\left(\frac{M_{1}}{10^8\rm g}\right)^{5/2}\left(\frac{ M_{2}}{10^{17}\rm g}\right)^{-1}\,.
     \label{betaratio}
\end{aligned} 
\end{equation}
Note that \eqref{betaratio} is independent of $w_A$ or $w_B$. The approximate equality sign corresponds to the case where PBH2 constitute for all CDM ($f_{\rm PBH}\equiv\rho_{\rm PBH2}/\rho_{\rm CDM}=1$). 

\begin{figure*}[ht]
    \centering
\includegraphics[width=.32\textwidth]{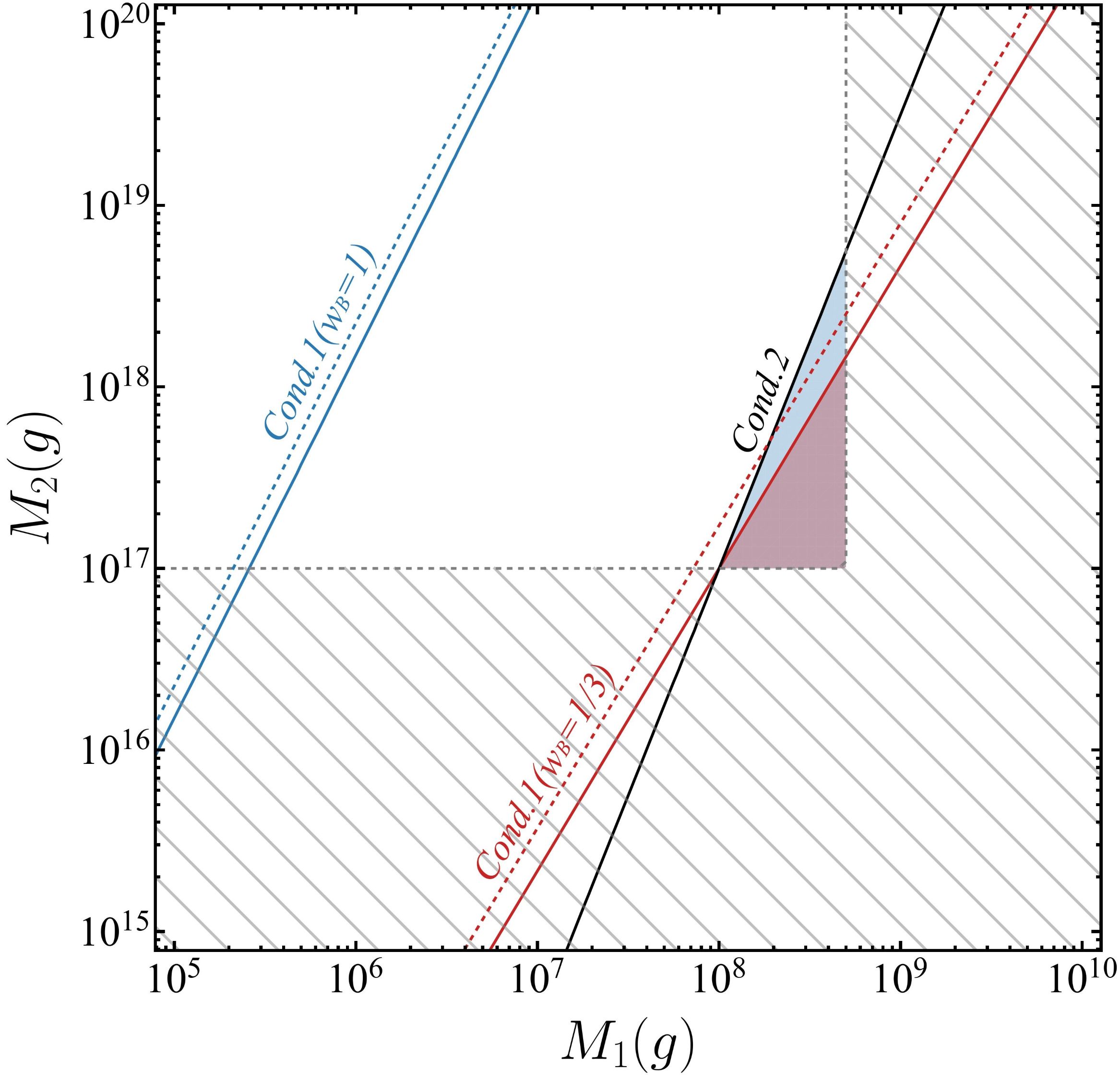}
\includegraphics[width=.32\textwidth]{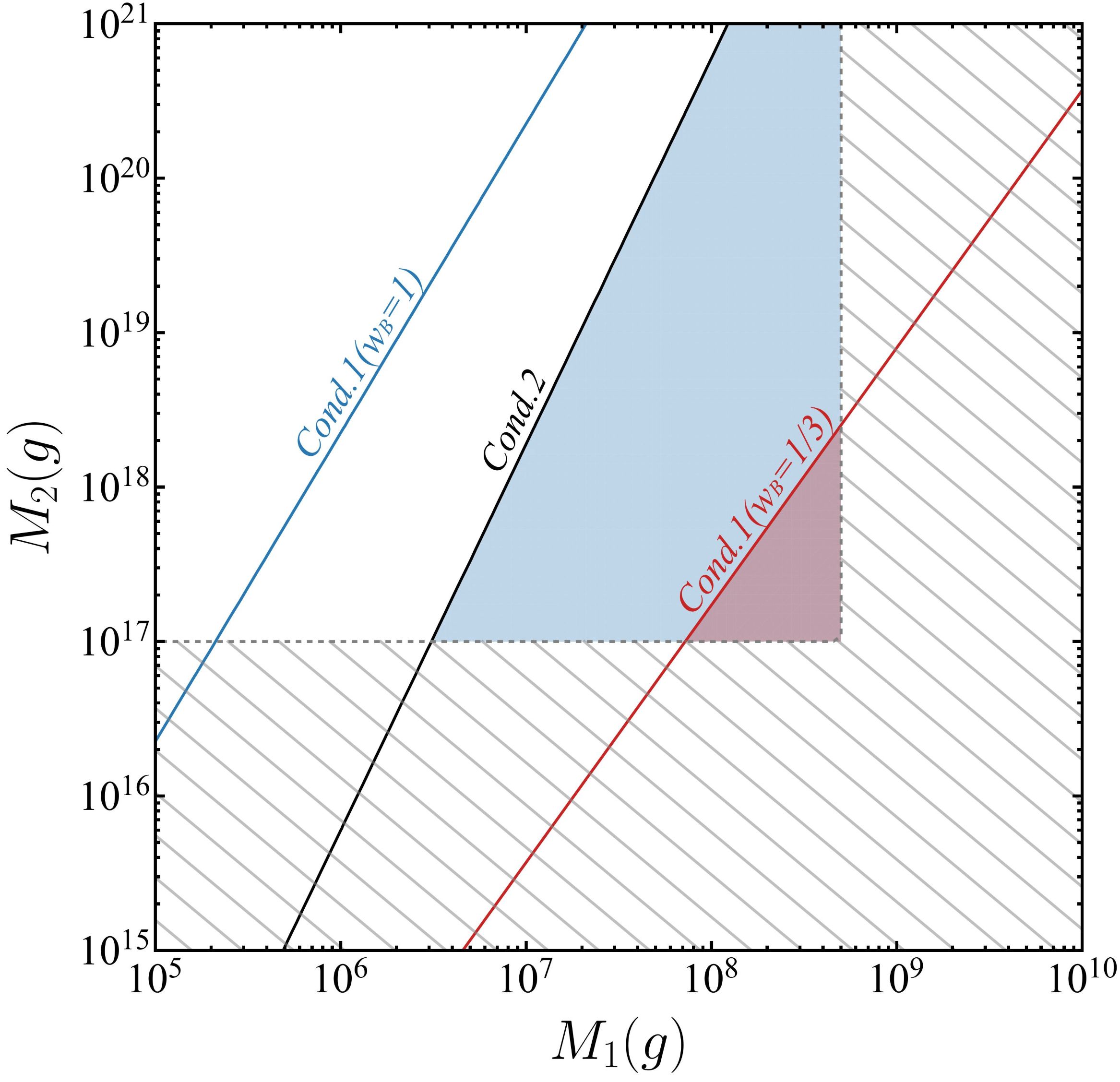}
\includegraphics[width=.32\textwidth]{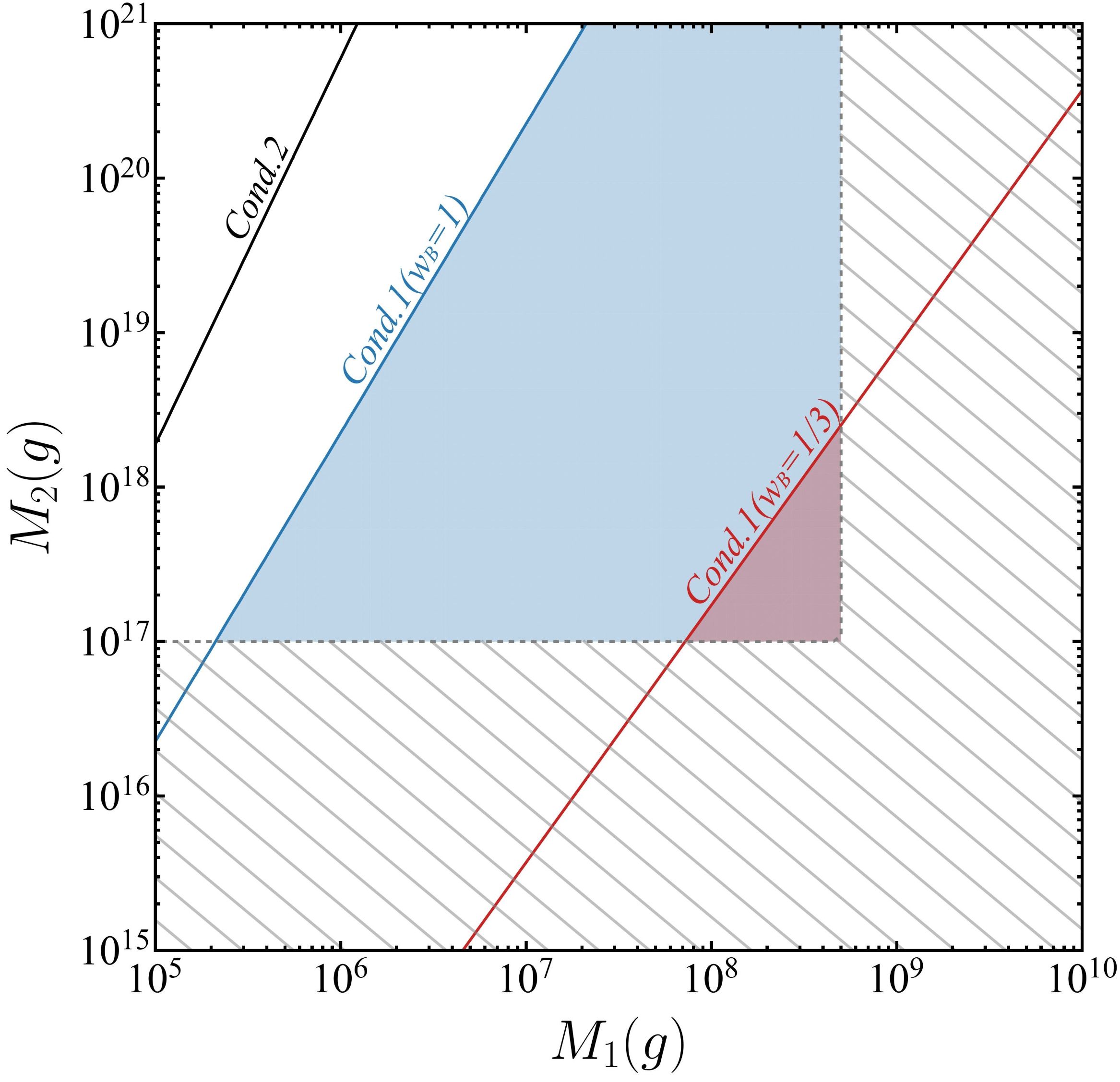}
\caption{\label{paracon} The parameter space for $m_1$ and $m_2$ that satisfies Condition~1~\eqref{con1} and Condition~2~\eqref{betaratio}.  
In the left figure, the solid lines for condition~1 are plotted take  $\beta_{1}=4.43\times 10^{-2}$,  $\beta_{2}=6\times 10^{-17}\beta_{1}^0$ and the dashed lines are plotted taking  $\beta_{1}=10^{-1}$,  $\beta_{2}=6\times 10^{-17}\beta_{1}$. In the middle figure, the solid lines for condition~1 are plotted taking $\beta_{1}=10^{-1}$,  $\beta_{2}=10^{-20}\beta_{1}$. In the right figure, the solid lines for condition~1 are plotted taking $\beta_{1}=10^{-1}$,  $\beta_{2}=10^{-25}\beta_{1}$.
The shaded areas are the parameter spaces for different cases, blue for $w_B=1$ and red for $w_B=1/3$.  The area filled with slashes is excluded to maintain the BBN bound for the mass of ultra-light PBHs $M_{1}<5\times 10^{9}\mathrm{g}$, and to respect the boundary of PBH dark matter window $10^{17}\mathrm{g}\lesssim M_2\lesssim 10^{21}\mathrm{g}$.}
\end{figure*}

In Fig.~\ref{paracon}, we plot the allowed region of parameter space satisfying \eqref{con1} and \eqref{betaratio} on the plane $(M_1,\,M_2)$ for different values of $\beta_1$.

A few comments are in order concerning Condition 2. It is evident from \eqref{betaratio} that the abundance of PBH2 must be much lower than that of PBH1 at the formation time. Since the abundance is sensitive to both the threshold~\footnotemark[3]\footnotetext[3]{Using Press-Schechter formalism, the PBH abundance at the formation for the Gaussian case is $\beta(M)=\gamma~\mathrm{erfc}\left({\delta_{\mathrm{th}}}/{\sqrt{2}\sigma_{\delta}(H)}\right)$, where ${\sigma_{\delta}^2(H)}\propto \sigma_{\mathcal{R}}^2(H)$ is the variance of the density contrast on scale $H$, smoothed by a window function $W(k; R)$.} and the amplitude of the curvature perturbation at horizon re-entry, 
there are three ways to realize the suppression of the PBH2 formation relative to the PBH1 formation: 
\begin{itemize}
    \item[(i)]  The curvature perturbation amplitude, which was once large before the first horizon re-entry $t_{\rm re,1}$, may be suppressed at the later stage due to the changes in the equation of state.
    \item[(ii)]  The curvature perturbation may become non-Gaussian after the transition from the break stage to the second inflationary stage, which may lead to the suppression of the probability density at the tail, hence of the PBH2 formation.
    \item[(iii)]  Due to the difference in the threshold values of the density perturbation for PBH formation. Namely, if $w_{B}>w_{A}$, the threshold at the $w_{B}$ stage is higher than that at the $w_{A}$ stage, i.e. $\delta_{\rm th,B}>\delta_{\rm th,A}$~\cite{Harada:2013epa}.
\end{itemize}

As the PBH formation rate is exponentially sensitive to the threshold value, the third mechanism, $w_{B}>w_{A}$, may be enough to realize our three-stage scenario, while the second one depends very much on the details of a model that realizes it. 
Hence, here we focus on the first mechanism to see if this can naturally account for the suppression of the PBH2 formation.

Assuming the curvature perturbation is Gaussian at both re-entries, using the Press-Schechter formalism, the PBH abundance is estimated as $\beta\propto{\rm erfc}[\delta_{\text{th}}/(\sqrt{2}\sigma)]$ where $\sigma^2=\int_0^\infty W^2(k,R)\mathcal{P}_\delta(t,k){\rm d}\ln k$ is the variance of the density contrast on the scale $R$, $W(k,R)$ is the window function that singles out the scale of interest, and $\mathcal{P}_\delta(t, k)$ is the density contrast power spectrum at the PBH formation, which is given by
\begin{align}\label{Pw}
    \mathcal{P}_{\delta}(t,k)\equiv\left(\frac{4}{9}\right)^2(kR)^4\mathcal{P}_{\mathcal{R}}(k), 
\end{align}
in the radiation domination, where $R$ is the comoving Hubble scale when PBHs form.
For simplicity, we consider a narrowly peaked curvature perturbation spectrum and ignore the window function in the following.
we assume radiation domination during both formation stages $w_A=w_B=1/3$, and consider a $\delta$-function peak spectrum $\mathcal{P}_\mathcal{R}(t,k)=\mathcal{A}(t)\delta(\ln[k/k_\star])$, which means $\sigma^2=16\mathcal{A}(t)/81$.
Then we obtain 
\begin{align}\label{ratiodelta}
  \frac{\beta_2}{\beta_1}\approx\frac{\mathcal{A}(t_{\rm re,2})}{\mathcal{A}(t_{\rm re,1})}\exp\left[-\frac{81}{32}\delta_{\rm th}^2\left(\mathcal{A}^{-1}(t_{\rm re,2})-\mathcal{A}^{-1}(t_{\rm re,1})\right)\right]\,.
\end{align}
Hence, for $\delta_{\rm th}\gtrsim0.4$, we find that in order to satisfy \eqref{betaratio} with $M_1\sim10^8{\rm g}$ and $M_2\sim10^{17}{\rm g}$, we should suppress the amplitude of the power spectrum at $t_{\rm re,2}$ to $\mathcal{A}(t_{\rm re,2})\lesssim10^{-2}$ provided that $\mathcal{A}(t_{\rm re,1})\simeq10^{-1}$.

\begin{figure}[htbp]
\includegraphics[width=.48
\textwidth]{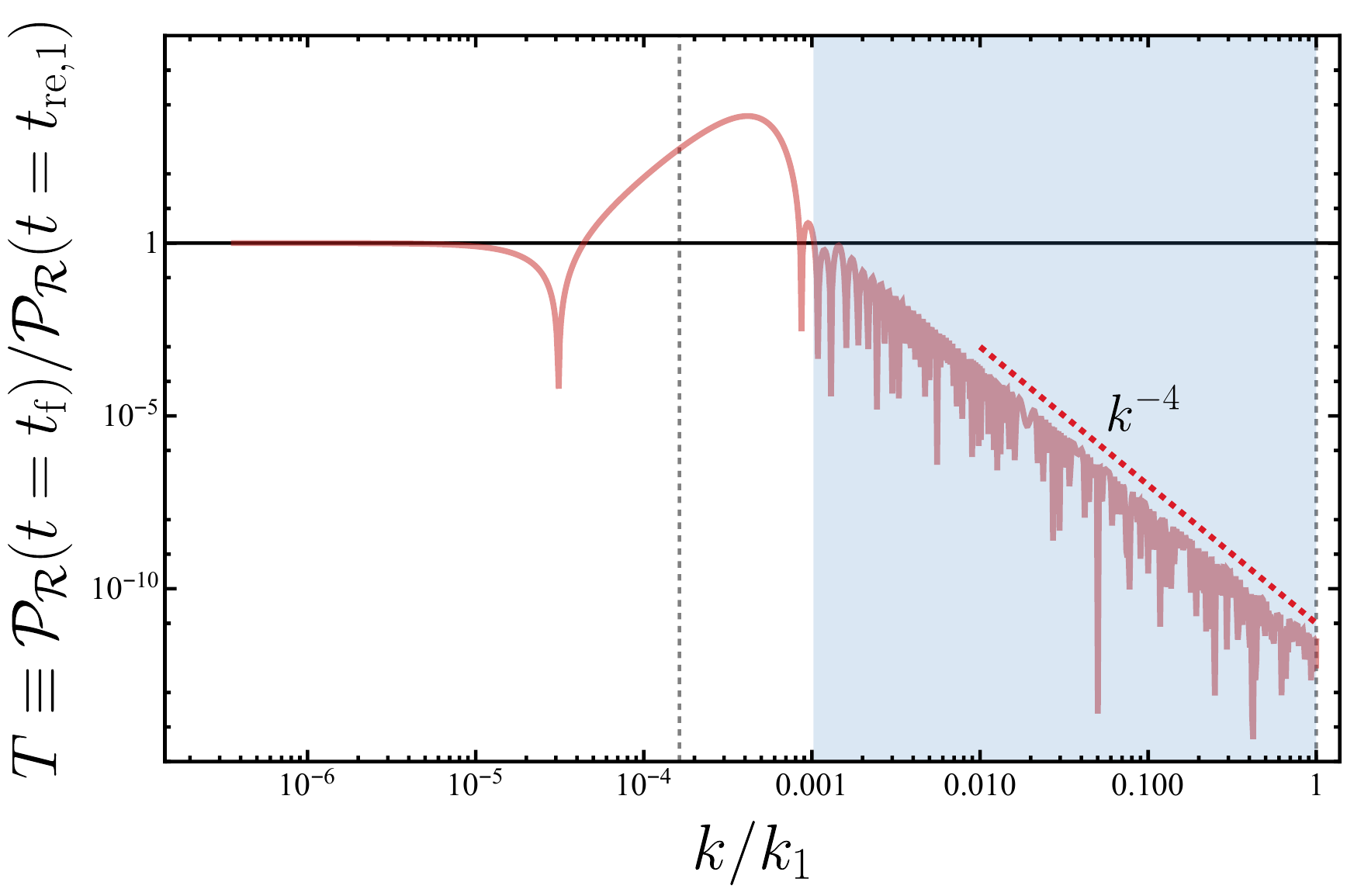}
\caption{\label{transferfunc}The transfer function $T$ for the power spectrum at the end of inflation $t_{\rm f}$ relative to the power spectrum at the beginning of break stage $t_1$, for smooth transitions of $w$ from $-0.999$ to $1/3$ and from $1/3$ to $-0.999$ with $c_s^2=1$ to $1/3$, and to $1$, with the duration of each transition being $O(1)$ e-folds.
The break ($w_A$) stage lasts about $\Delta N=9$ e-folds.  
The vertical dotted line denotes $k=k_2$. 
The light blue area shows the region of suppression. As shown in the figure, in the region of suppression, the power spectrum follows the scaling behavior $k^{-4}$.
}
\end{figure}

In order to estimate the suppression factor, we numerically solve the evolution of the comoving curvature perturbation, 
\begin{align}
    \mathcal{R}''_{k}+2\frac{z'}{z}\mathcal{R}'_{k}+c_s^2k^2\mathcal{R}_{k}=0,
    \label{req}
\end{align}
where  $z\equiv a\sqrt{3(1+w)}M_{\rm pl}/c_s$ and $c_s$ is the sound velocity.
We approximate the equations of state at the two stages of inflation by $w=-1$ and at the break stage by $1/3$, and numerically calculate the transfer function $T\equiv\mathcal{P}_\mathcal{R}(k, t_{\rm f})/\mathcal{P}_\mathcal{R}(k, t_{\rm re,1})$, i.e., the ratio of the power spectrum at end of inflation $t_{\rm f}$ to that at the first horizon re-entry $t_{\rm re,1}$. 
As shown in Fig.~\ref{transferfunc}, in the blue shaded region, the amplitude of the curvature perturbation at wavenumber $k_\star\gtrsim0.1k_1$ is suppressed as $\propto k^{-4}$, which can easily give more than an order of magnitude suppression at the second re-entry. 
This implies that one can obtain a sufficient suppression to satisfy~\eqref{betaratio} for an original spectrum sharply peaked at $k=k_\star\gtrsim0.1k_1$.



In the above, we assumed that PBH1 and PBH2 form at the same comoving wavenumber.
This may not be the case if the original power spectrum spreads over a finite width. In this case, the abundance of PBH1 and PBH2 may be characterized by different wavenumbers.

If $k_2\ll k_1$, one may consider the case when most of PBH1 form at the beginning of the break stage ($t\gtrsim t_1$) with comoving wavenumber $k\lesssim k_1\equiv a(t_1)H_{\rm I}$. 
Then, if the amplitude of the curvature perturbation at $k\ll k_1$ is suppressed at the end of inflation, a peak appears at $k_{\star2}\gtrsim k_2\equiv a(t_2)H_{\rm II}$.

\begin{figure}[htbp]
\includegraphics[width=.48
\textwidth]{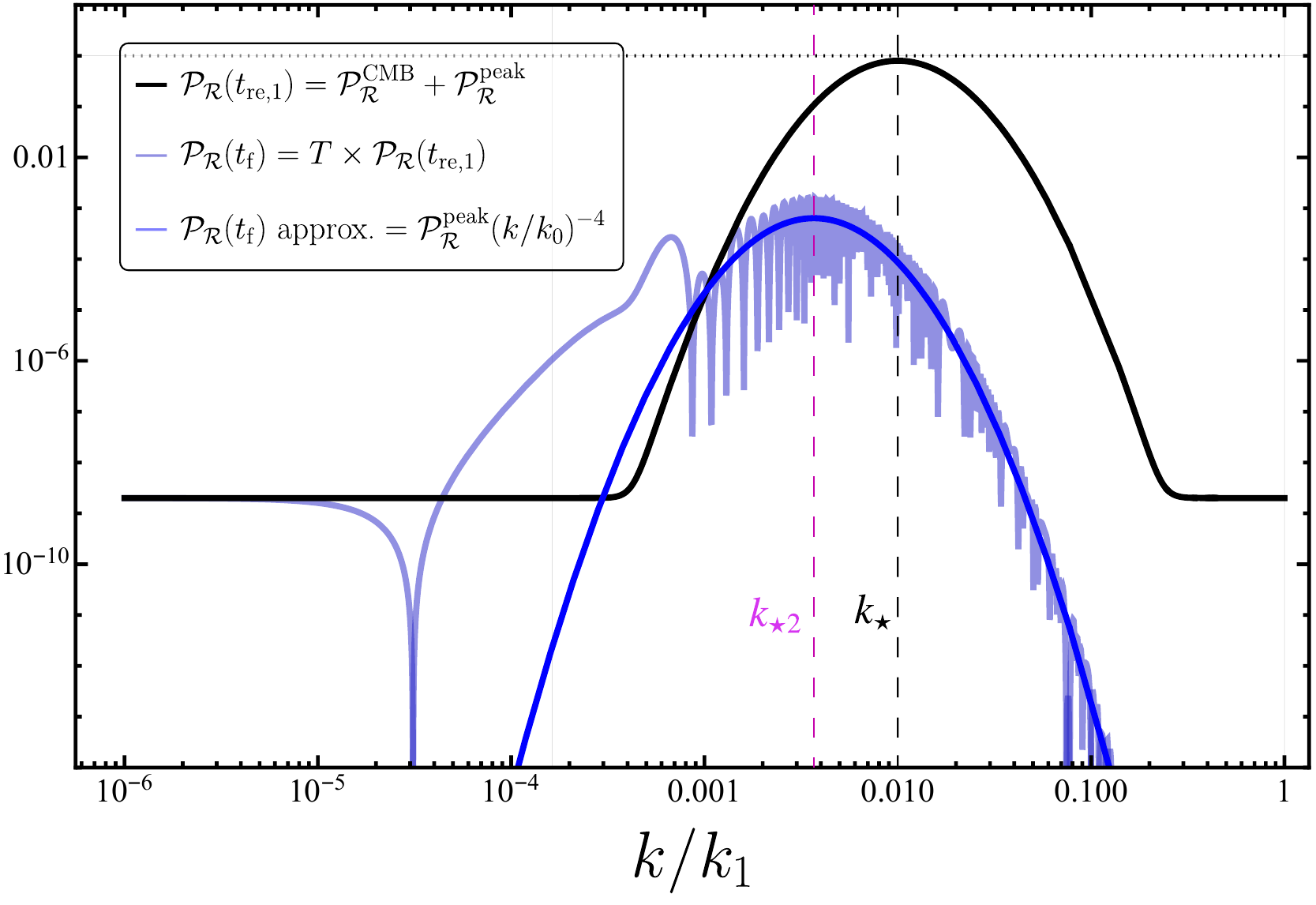}
\caption{\label{prsuppression} The curvature perturbation power spectra measured at the beginning of break stage $t_1$ and the end of inflation $t_\mathrm{f}$. 
Smooth transitions are considered where $w=-0.999$ and $1/3$ during the two inflationary stages and the break stage, respectively, with $\Delta N=9$. 
The black curve is the initial power spectrum, which is the sum of a scale-invariant part $\mathcal{P}_{\mathcal{R}}^{\mathrm{CMB}}=2\times 10^{-9}$ and a lognormal peak part $\mathcal{P}_{\mathcal{R}}^{\mathrm{peak}}=\frac{\mathcal{A}_{\mathcal{R}}}{\sqrt{2\pi}\Delta}\exp\left(-\frac{\ln^2(k/k_\star)}{2\Delta^2}\right)$ with  $\mathcal{A}_{\mathcal{R}}=1$, $\Delta=0.5$ and $k_\star=0.01k_1$.
The purple curve is the spectrum at the end of inflation $\mathcal{P}_{\mathcal{R}}(t_{\mathrm{f}})$. 
The blue curve is an approximation given by $\mathcal{P}_{\mathcal{R}}(t_{\mathrm{f}})\approx\mathcal{P}_{\mathcal{R}}^{\mathrm{peak}}(k)(k/k_0)^{-4}$ where
$k_0$ is the point at which the transfer function $T$ is equal to unity (see Fig.~\ref{transferfunc}).
}

\end{figure}

In order to check if the suppression factor can satisfy \eqref{betaratio}, we consider an enhanced power spectrum at the first horizon reentry $t_{\rm re,1}$ as an example; $\mathcal{P}_\mathcal{R}(t_{\rm re,1})=\mathcal{P}_\mathcal{R}^{\rm CMB}+\mathcal{P}_{\mathcal{R}}^{\mathrm{peak}}$ where we parameterize the enhanced spectrum by the lognormal function peaked at $k_\star$ $(k_2\leq k_\star\leq k_1)$,
\begin{align}
\mathcal{P}_{\mathcal{R}}^{\mathrm{peak}}(k)\equiv\frac{\mathcal{A}_{\mathcal{R}}}{\sqrt{2\pi}\Delta}\exp\left(-\frac{\ln^2(k/k_\star)}{2\Delta^2}\right).
\label{lognormalpeak}
\end{align}
Using the transfer function $T$ calculated previously, we find $\mathcal{P}_\mathcal{R}(t_{\rm f},k)$, shown in Fig.~\ref{prsuppression}.
We parameterize the peak part of the spectrum at the end of inflation as $\mathcal{P}_\mathcal{R}(k;t_{\rm f})=\mathcal{P}_\mathcal{R}^{\rm peak}(k)(k/k_0)^{-4}$ according to the scaling behavior shown in Fig.~\ref{transferfunc}, where $k_0$ is at which the transfer function crosses unity $T(k_0)=1$ within the range $k_2\leq k\leq k_1$.  

From \eqref{lognormalpeak}, we obtain the variance of the density contrast $\sigma^2$ defined beyond \eqref{Pw} at the formation of PBH1 for the peak comoving wave number $k_R=k_\star$. (Note that since we are focusing on the sharp peak case, we ignore the window function.)
\begin{align}
    &\sigma^2 (M_{1})=\frac{16}{81}\mathcal{A}_{\mathcal{R}}e^{8\Delta^2}
\end{align}
Note that this matches the $\delta$-function peak case in the limit $\Delta\to0$. 
For a narrow spectrum $\Delta\lesssim1$, we find that $\mathcal{P}_\mathcal{R}(k;t_{\rm f})=\mathcal{P}_\mathcal{R}^{\rm peak}(k) (k/k_0)^{-4}$ is peaked at $k_{\star2}=e^{-4\Delta^2}k_\star$. 
Hence, we may approximate it by a lognormal spectrum with the peak value $\mathcal{A}_{\mathcal{R}}(k_0/k_{\star2})^4$ at $k=k_{\star2}$. 
This gives the variance for the PBH2 formation as
\begin{align}
     \sigma^2 (M_{2})\approx\frac{16}{81}\mathcal{A}_{\mathcal{R}}e^{8\Delta^2}\left(\frac{k_0}{k_{\star2}}\right)^{4}.
\end{align}
Requiring $\sigma^2(M_2)<\sigma^2(M_1)$ implies the condition $k_0<k_{\star2}$.
Thus, for a narrow spectrum $\Delta\lesssim1$ and for $k_2\ll k_1$, it is always possible to suppress the PBH2 abundance by selecting $k_0<k_{\star2}\lesssim k_\star$.

Applying a similar argument used to derive \eqref{ratiodelta}, we obtain the ratio of the PBH2 to PBH1 abundances for a finite-width spectrum as
\begin{widetext}
\begin{equation}
\begin{aligned}
     & \frac{\beta_2}{\beta_1}\approx\left(\frac{k_0}{k_\star}\right)^2\exp\Biggl[4\Delta^2
     +\frac{81}{32}e^{-16\Delta^2}\frac{\delta_{\text{th}}^2}{\mathcal{A_R}}\left({e^{8\Delta^2}}-\left(\frac{k_\star}{k_0}\right)^4\right)\Biggr].
\end{aligned}
\label{ratiowidth}
\end{equation}
\end{widetext}
Let us take $\mathcal{A_{R}}=1$, $\Delta=0.5$ and $\delta_{\text{th}}= 0.4$ as a specific case. Then, Condition 2 \eqref{betaratio} with $M_1\sim10^8{\rm g}$ and $M_2\sim10^{17}{\rm g}$ is satisfied for $k_\star>8.24k_0$.
Therefore, either in the case of a delta-function peak or with finite width, if the amplitude of the peak at $t_{\rm f}$ is at least one order lower than that at $t_{\rm re,1}$, the condition \eqref{betaratio} can be satisfied.

\section{\label{sec:level1} Gravitational Wave signatures}
In this section, we evaluate the induced GWs (IGWs) emitted during the evaporation of PBH1 and during the PBH2 formation era.

\subsubsection{GWs induced by PBH isocurvature}
As shown in Fig.~\ref{rhopbh}, we assume that PBH2 are always subdominant to PBH1 before they evaporate, i.e, $\rho_{\rm PBH2}\ll\rho_{\rm PBH1}$. 
Assuming the Poisson distribution of PBHs, the ultra-violet (UV) cut-off of the PBH1 density spectrum is given by
\begin{align}
    k_{\rm UV}\equiv\frac{a_{\rm re,1}}{d}
    =k_\star\left(\frac{\beta_{1}}{\gamma}\right)^{1/3}\,(\ll k_\star),
    \label{kkstar}
\end{align}
where $d$ is the mean physical separation of PBH1.
Before the PBH1 domination, the density fluctuations of PBH1 are isocurvature perturbations, 
\begin{align}
    S=\frac{\delta\rho_{\rm PBH1}}{\rho_{\rm PBH1}}-\frac{\delta\rho_{B}}{(1+w_B)\rho_{B}}\,.
\end{align}
Neglecting the curvature perturbation on large scales, $k\gg k_\star$, we have $\delta{\rho_B}+\delta{\rho_{\rm PBH1}}\approx0$ and $\rho_B\gg\rho_{\rm PBH1}$.
Hence $S\approx{\delta\rho_{\rm PBH1}}/{\rho_{\rm PBH1}}$, and the initial power spectrum of
$S$ is given by
\begin{align}
    \mathcal{P}_{S}(k)=\frac{2}{3\pi}\left(\frac{k}{k_{\rm UV}}\right)^3.
\end{align}
After PBH-domination, the isocurvature perturbations convert into adiabatic perturbations and source GWs~\cite{Domenech:2023jve,Inomata:2019ivs,Domenech:2020ssp,Inomata:2020lmk}. 
According to \cite{Domenech:2021ztg,Domenech:2024wao}, the peak frequency of the GW spectrum today is related to the UV cut-off scale given in the isocurvature power spectrum,
\begin{align}
    f_{\rm UV}&=\frac{k_{\rm UV}}{2\pi a_0}\approx 2.06\times10^{-1} \mathrm{Hz}\left(\frac{M_1}{5\times10^8\rm g}\right)^{-5/6}\notag\\
    &\qquad\times\left(\frac{g_\star(T_{\rm eva})}{106.75}\right)^{1/4}\left(\frac{g_{\star,s}(T_{\rm eva})}{106.75}\right)^{-1/3}.
\end{align}
based on entropy conservation
\footnotemark[4]\footnotetext[4]{For radiation at temperature $T$, the entropy can be calculated by $s=\frac{2\pi^2}{45}g_{\star s}(T)T^3$, for $T>100\mathrm{GeV}$ we have $g_{\star s}\approx g_{\star}\approx 106.75$. While at the time of matter-radiation domination$g_{\star s}(T_\mathrm{eq})\approx g_{\star s}(T_0)\approx 3.94$, $g_{\star }(T_\mathrm{eq})\approx g_{\star }(T_0)\approx 3.38$. Considering that the total entropy is conserved, which means $s\propto a^{-3}$, we should obtain $a/a_0=(T_0/T)(g_{\star s}(T_0)/g_{\star s}(T))^{1/3}$, combining with the scaling between the Hubble $H(T)$ and temperature $H(T)\propto T^{3(1+w)/2}$, as well as the relation between Hubble and the PBH mass formed (\eqref{mpbhf}), we could obtain the frequency based on the thermal history of the universe\cite{Domenech:2021ztg}. In our case, it's simple to first derive the $f_{\mathrm{eva}}$ based on the standard thermal history, and then transfer to the $f_{\mathrm{UV}}$ and $f_\star$ using their relation \eqref{kkstar} and (2.16) from \cite{Domenech:2024wao}}. 
For $M_{1}\approx 5\times 10^8 \rm g$, $f_{\rm UV}\sim 10^{-1}{\rm Hz}$ lies in the LISA/Taiji/TianQin/DECIGO band.
The amplitude of the spectral peak today is estimated as
\begin{align}
    &\Omega_{\mathrm{GW}}^{\mathrm{eva}}h^2( k_\mathrm{UV})\approx\mathcal{F}\left(\frac{k_\mathrm{UV}}{k_{\mathrm{eva}}}\right)^{17/3}\left(\frac{k_{\mathrm{eq}}}{k_\mathrm{UV}}\right)^8\nonumber\\ &\approx\mathcal{F}\left(\frac{2\pi\gamma }{\alpha}\right)^{17/9}\left(\frac{M_{\mathrm{1}}}{M_{\mathrm{2}}}\right)^{\frac{8(n_B-1)}{2n_B-3}}\left(\frac{M_{\mathrm{1}}}{M_{\mathrm{pl}}}\right)^{34/9}\beta_{1}^{\frac{8n_B}{3(2n_B-3)}},
\end{align}
where
\begin{align}   
\mathcal{F}=3\times10^{-10}\left(\frac{\Omega_{r,0}h^2
}{4.18\times10^{-5}}\right)\left(\frac{g_*(f)}{106.75}\right)\left(\frac{g_{*,s}(f)}{106.75}\right)^{-4/3}. 
\end{align}


\subsubsection{GWs induced by enhanced curvature perturbations}
In addition to GWs produced at PBH1 evaporation, the enhanced curvature perturbation at $k_{\star}$ can induce GWs at second order \cite{Matarrese:1992rp, Matarrese:1993zf,Nakamura:1996da, Matarrese:1997ay, Noh:2004bc, Carbone:2004iv, Nakamura:2004rm, Ananda:2006af, Osano:2006ew}. 
Here we note that PBH2s are produced during the $w_B$ stage, not during radiation dominance after evaporation of PBH1.
The frequency of the spectral peak today is given by 
\begin{equation}
    \begin{aligned}
    f_{\star}&=\frac{k_{\star}}{2\pi a_0}= \left(\frac{\gamma}{\beta_{1}}\right)^{1/3}f_{\rm UV}\\&\approx5.6\times 10^{-1} \mathrm{Hz} \left(\frac{\gamma}{0.2}\right)^{1/3} \left(\frac{\beta_{1}}{10^{-2}}\right)^{-1/3}\\
    & \times\left(\frac{M_1}{5\times 10^8\rm g}\right)^{-5/6}\left(\frac{g_\star(T_{\rm eva})}{106.75}\right)^{1/4}\left(\frac{g_{\star,s}(T_{\rm eva})}{106.75}\right)^{-1/3}\,.
\end{aligned}
\end{equation}
For $M_{1}=5\times 10^8\mathrm g$ and $\beta_{1}=10^{-2}$, we have $f_{\star}\approx 7.06\times 10^{-1}\mathrm{Hz}$, which lies in the DECIGO/ET/LVK band. 
Since the GWs induced by enhanced curvature perturbation at $t_{\rm re,1}$ are significantly diluted during the second inflationary stage, in the following, we  mainly consider the GWs induced at $t_{\rm re,2}$. 
The ratio of the energy density at matter-radiation equality $t_{\rm m}$ to that at the time of production is 
\begin{equation}
    \begin{aligned}
&\frac{\left.\Omega_{\mathrm{GW}}^{\mathcal {R}}\right|_{t=t_{\rm m}}}{\left.\Omega_{\mathrm{GW}}^{\mathcal {R}}\right|_{t=t_{\rm re,2 }}}=\left(\frac{a_{\rm{eq}}}{a_{\rm{re,2}}}\right)^{2n_B-4}\left(\frac{a_{\rm{eva}}}{a_{\rm{eq}}}\right)^{-1}\\&=2.53\left(\frac{\beta_{1}}{ 10^{-2}}\right)^{-4/3}\left(\frac{M_2}{10^{17} \mathrm{g}}\right)^{2}\left(\frac{M_1}{5\times 10^{8} \mathrm{g}}\right)^{-10/3}\,.
\end{aligned}
\end{equation}
Interestingly, this ratio depends on both the mass and abundance of the formed PBHs but not on the EOS parameter $w_B$. 

\begin{figure}[htbp]
\includegraphics[width=.48
\textwidth]{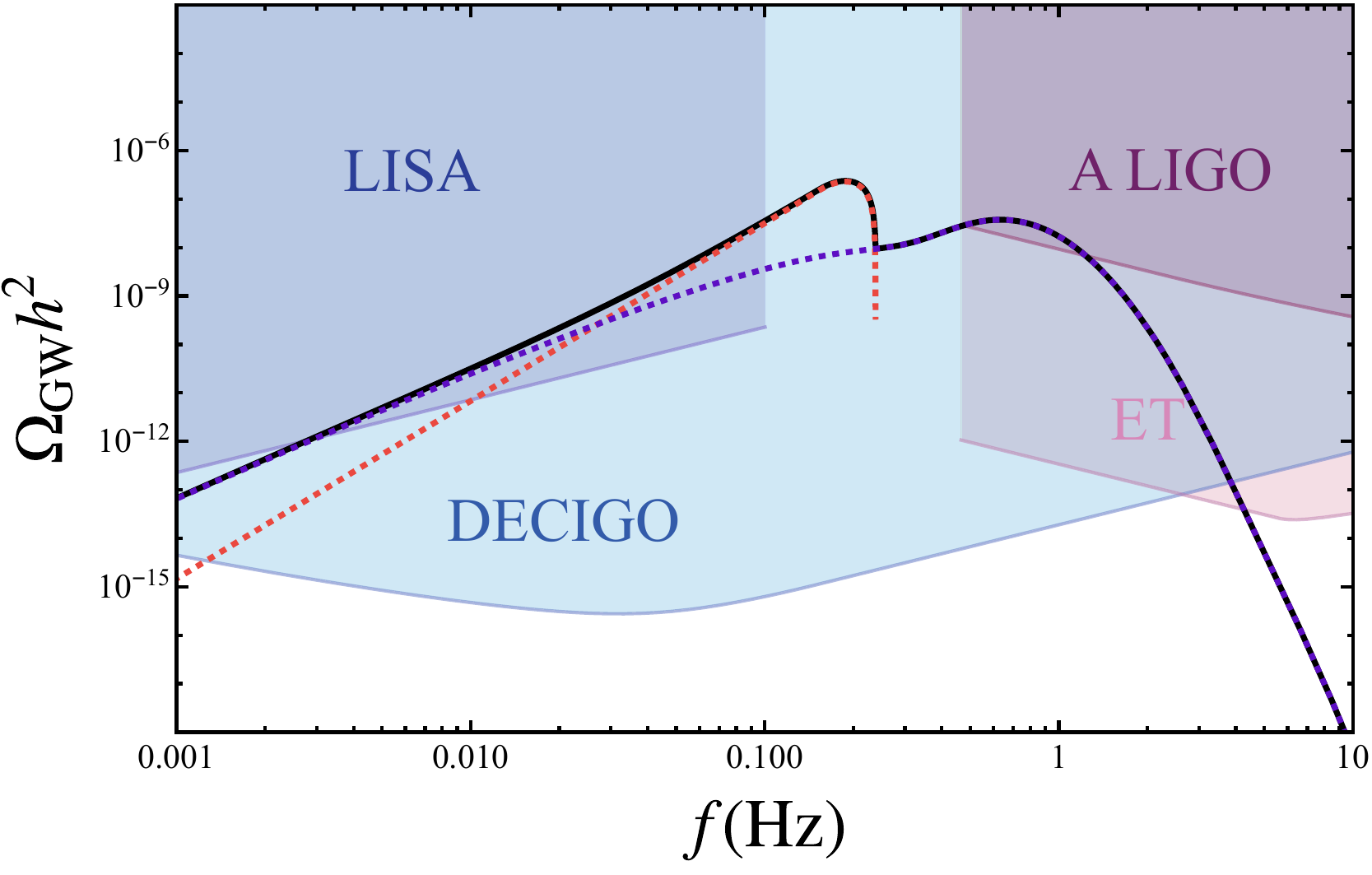}
\caption{\label{gwformation}
The induced GW spectrum for the dual PBH scenario, for $M_{\mathrm{PBH,1}}=5\times 10^8 \mathrm{g}$ , $M_{\mathrm{PBH,2}}=10^{17} \mathrm{g}$, $\beta_1=10^{-2}$, 
 and $A_\mathcal{R}=0.05$. 
 The red dashed curve is the GW spectrum from PBH1 evaporation, and the purple dashed curve is from PBH2 formation. The black solid curve is the total GW spectrum. 
 We also plot the sensitivity curve of the gravitational wave detectors \cite{Thrane:2013oya,Schmitz:2020syl}, including  LISA~\cite{LISA:2022yao}, DECIGO~\cite{Seto:2001qf}, LIGO A+~\cite{APlusDesignCurve}, and ET~\cite{ET:2019dnz}.}
\end{figure}

For definiteness, following the assumption of the lognormal form for the initial curvature perturbation spectrum \eqref{lognormalpeak}, we approximate the spectrum at the second reentry $t_{\rm re,2}$ by another lognormal form, $\mathcal{P}_{\mathcal{R}}(k;t=t_{\mathrm{re},2})=\frac{\mathcal{A}_{\mathcal{R}2}}{\sqrt{2\pi}\Delta}\exp\left(-\frac{\ln^2(k/k_{\star2})}{2\Delta^2}\right)$, where $\mathcal{A}_{\mathcal{R}2}=\mathcal{A}_{\mathcal{R}}(k_0/k_{\star2})^4$. 
Then the peak amplitude of the GW spectrum today is expressed as~\cite{Pi:2020otn}
\begin{align}
 &\Omega_{\mathrm{GW},0}^{\mathcal{R}}h^2(k= k_{\star})\approx5.6\times10^{-6}\mathcal{A}_{\mathcal{R}2}^2\,
       \frac{\left.\Omega_{\mathrm{GW}}^{\mathcal {R}}\right|_{t=t_{\rm m}}}{\left.\Omega_{\mathrm{GW}}^{\mathcal {R}}\right|_{t=t_{\rm re,2 }}}
       \nonumber\\
       &\quad\times\left(\frac{\Omega_{r,0}h^2
}{4.18\times10^{-5}}\right)\left(\frac{g_*(f)}{106.75}\right)\left(\frac{g_{*,s}(f)}{106.75}\right)^{-4/3},
\end{align}
where we have adopted $\Delta=0.5$. 
In Fig.~\ref{gwformation}, we plot the total power spectrum, $\Omega_{\rm GW}h^2=\Omega_{\rm GW}^{\rm eva}h^2+\Omega_{\rm GW}^{\mathcal R}h^2$, where PBH1s are assumed to be light enough to evaporate away before BBN, while PBH2s are sufficiently heavy and abundant to explain the origin of CDM today. 

A characteristic feature of the GW spectrum in the dual PBH scenario is that there appear two peaks, originating from the isocurvature induced GWs from the evaporation of PBH1 (red dashed curve in Fig.~\ref{gwformation}), and the curvature induced GWs at the second horizon reentry (purple dashed curve), respectively. 
Both of them are within the range of detectability of the current or future GW detectors.


\section{\label{sec:level1} Discussion and Conclusion}
In this letter, we proposed the dual PBH formation scenario in two-stage inflation with a break. 
Light PBHs with mass $\lesssim\mathcal{O}(10^8{\rm g})$ (PBH1s) form when the enhanced primordial curvature perturbations first re-enter the Hubble horizon during the break stage. 
After inflation, they lead to a PBH-dominated stage, followed by a radiation-dominated stage from their evaporation. 
On the other hand, PBHs with mass $\gtrsim\mathcal{O}(10^{17}{\rm g})$ (PBH2) can form at the second re-entry into the Hubble horizon after inflation, hence may serve as a candidate for CDM. 
Thus, the dual PBH scenario naturally explains reheating of the universe which lead to the radiation-dominated stage, while explaining the origin of CDM at the same time.

Interestingly, our scenario predicts a bi-peak signature in the GW power spectrum, one from the evaporation of lighter PBH1, and the other from the epoch of heavier PBH2 formation. 
It is falsifiable because the amplitudes and frequencies of both peaks are located within the detectable range of future GW observatories such as LISA, DECIGO, Advanced LIGO, and ET.  

Besides the bi-peak feature, there may appear other features in the GW power spectrum. For example, non-Gaussianities in the primordial curvature perturbations may source the clustering of PBH1s before their evaporation~\cite{Suyama:2019cst,Domenech:2020ssp}, hence leading to another peak at low frequency in the GW spectrum~\cite{Papanikolaou:2024kjb, He:2024luf}. Moreover, recently it has been pointed out that the primordially adiabatic mode may also induce non-negligible GW signatures~\cite{Domenech:2024wao}. The investigation of these observational issues in the dual PBH scenario are left for future study.

\begin{acknowledgments}
We thank Kazunori Kohri, Kaloian Lozanov, Xiao-Han Ma, Shi Pi, Jan Tränkle, Tsutomu Yanagida for useful discussions. 
Kavli IPMU is supported by World Premier International Research Center Initiative (WPI), MEXT, Japan.
This work is supported in part by JSPS KAKENHI Grant Nos. JP20H05853 and JP24K00624.
X.W. is supported by Forefront Physics and Mathematics Program to Drive Transformation (FoPM), a World-leading Innovative Graduate Study (WINGS) Program, the University of Tokyo.
Y.Z. is supported by  the Fundamental Research Funds for the Central Universities, and by the Project 12475060 and 12047503 supported by NSFC, Project 24ZR1472400 sponsored by Natural
Science Foundation of Shanghai, and Shanghai Pujiang Program 24PJA134. 
X.W. and Y.Z. gratefully acknowledge the hospitality and support of the Kavli Institute of Physics and Mathematics of the Universe (Kavli IPMU), the University of Tokyo during their visit when the work is done. 

We would also like to acknowledge the valuable input and discussions from the workshop "new perspectives on cosmology" held at APCTP, Korea, which greatly contributed to the completion of this work.

\end{acknowledgments}
\appendix
\section{About PBH2 formation time}
Here, we prove that the PBH2 formation must happen before the moment PBH1 starts to dominate the universe. 
Assuming PBH2 forms during the PBH1 dominated era, the equality \eqref{hubeq} is converted to 
\begin{align}
   \frac{a_{\mathrm{re,2}}}{a_\mathrm{re,1}}=\left(\frac{a_{2}}{a_{\mathrm{re,1}}}\right)^{A}\left(\frac{a_\mathrm{eq}}{a_{\mathrm{f}}}\right)^B\left(\frac{a_{\mathrm{re,2}}}{a_\mathrm{eq}}\right)^{3/2}
\end{align}
Inserting \eqref{abund1} into the upper equality, we have
\begin{align}
     \frac{a_{\mathrm{re,2}}}{a_\mathrm{re,1}}=\beta_{1}<1\,,
\end{align}
which is inconsistent with our assumption that $a_{\mathrm{re,2}}>{a_\mathrm{re,1}}$. 
Therefore, we conclude that PBH2 should always form before the PBH1 dominated stage. 
Another intuitive way to show this is as follows.
Since $\beta_1<1$, the mean comoving separation (or the UV cut-off scale) of PBH1 should be larger than the Hubble horizon when they formed, hence $k_{\mathrm{UV}}<k_{*}$. 
At the onset of the PBH1 domination, the mean separation should be smaller than the Hubble horizon so that there is at least one PBH within the horizon, which means $k_{\mathrm{eq}}<k_\mathrm{UV}$. 
As a result, we have $k_{\mathrm{eq}}<k_\star$.
\vfill

\bibliography{apssamp}
\bibliographystyle{apsrev4-1}
\end{document}